\documentclass[a4paper, 10 pt, conference]{hsmr} 
\usepackage[utf8]{inputenc}
\IEEEoverridecommandlockouts                       
\usepackage{mathtools}
\usepackage{geometry}
\usepackage[labelsep=space]{caption}
\usepackage{newtxmath,newtxtext}
\usepackage{authblk}
\usepackage{pgfplots}
\usepackage{graphicx}
\usepackage{gensymb}
\usepackage[colorlinks,urlcolor=blue]{hyperref} 
\usepackage{newtxtext,newtxmath}

\pgfplotsset{compat=newest}

\graphicspath{{./images/}}
\DeclareGraphicsExtensions{.pdf,.png,.jpg}

\makeatletter
\let\NAT@parse\undefined
\makeatother
\usepackage[numbers,compress]{natbib}
\bibpunct{[}{]}{,}{n}{}{;}
\definecolor{orange}{rgb}{1,0.2,0}

\usepackage{subfig}

\title{
\LARGE \bf
Guidance for Intra-cardiac Echocardiography Manipulation to Maintain Continuous Therapy Device Tip Visibility
} 

\author{\large Jaeyoung Huh$^{}$}
\author{\large Ankur Kapoor$^{}$} 
\author{\large Young-Ho Kim$^{}$}

\affil{\normalsize\textit{$^{}$Digital Technology \& Innovation, Siemens Healthineers, Princeton, NJ, USA,} \\ 
\small\textit{(jaeyoung.huh, ankur.kapoor, young-ho.kim)@siemens-healthineers.com}}

\begin{document}

\maketitle
\thispagestyle{empty}
\pagestyle{empty}

\section*{INTRODUCTION}

Intra-cardiac Echocardiography (ICE) is a key imaging modality for Electrophysiology (EP) procedures and Structural Heart Disease (SHD) interventions, providing real-time, high-resolution visualization of intracardiac structures. In EP procedures, accurate ablation catheter tip tracking is essential for precise lesion formation, while in SHD interventions, ICE aids in positioning structural therapy devices such as mitral clips, occluders, and valve implants.

However, manual ICE catheter manipulation presents challenges due to limited direct visualization and frequent adjustments, making consistent and accurate device tracking difficult. To address this, robot-assisted ICE catheter control has been explored to enhance accuracy and reduce operator workload,\cite{kim2022automated,huh2025viewguidance, kim2020automatic}. For effective robotic assistance, accurately estimating the direction and angle of devices as they appear in ICE images is critical in guiding the automated movement of the ICE catheter.

This study proposes an AI-driven model that estimates the incident angle of therapy devices within the 2D ICE imaging field. We focus on predicting the ablation catheter tip direction as a tip heading incident vector in real time. To evaluate performance, we analyze the necessary ICE catheter adjustments to maintain continuous device visibility, providing a foundation for future robot-assisted ICE catheter control.



\section*{MATERIALS AND METHODS}
\begin{figure}[t!]
	\centering 
	\includegraphics[width= 0.47\textwidth]{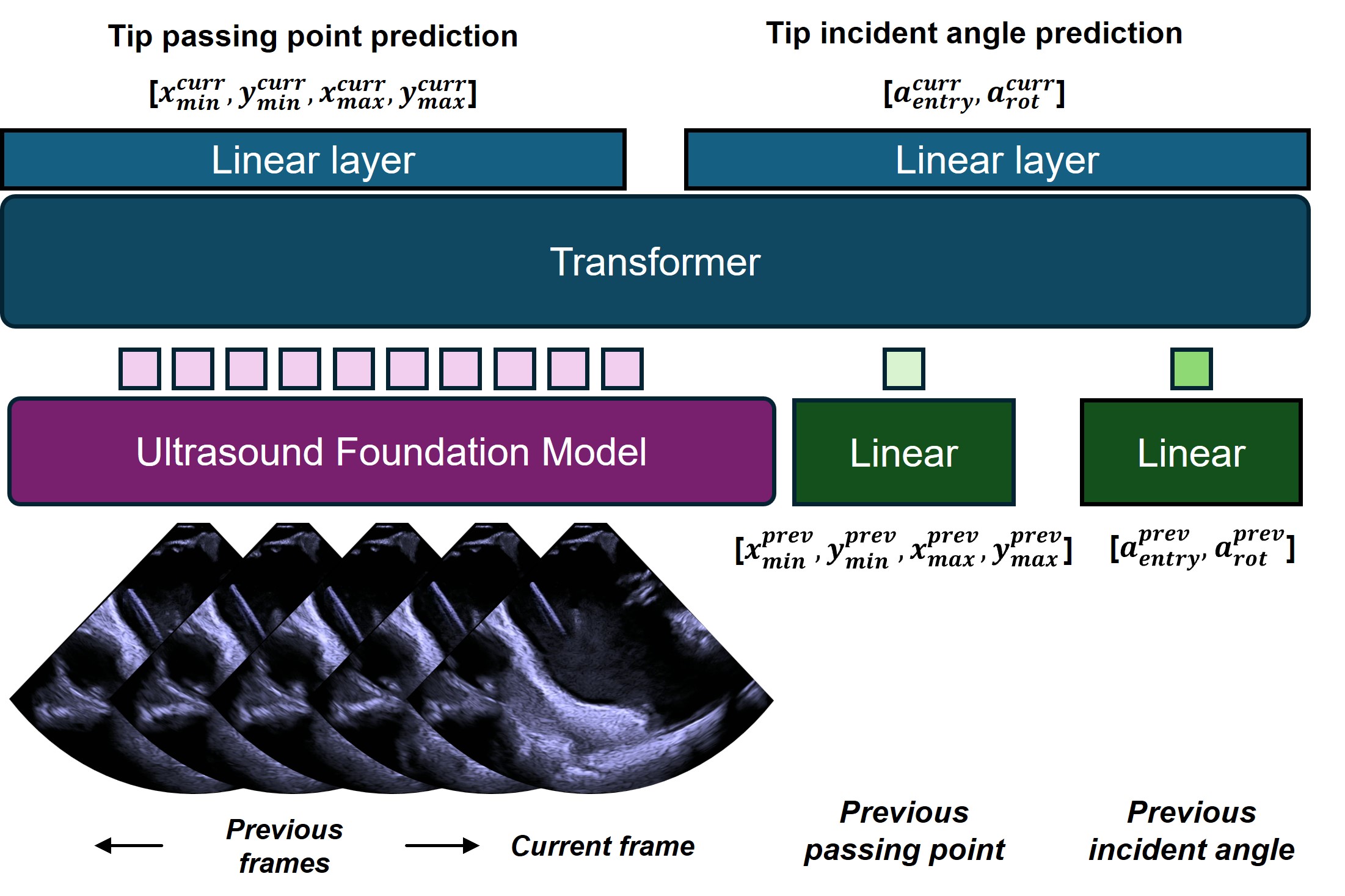}
	\caption{A series of ICE images are processed by the US foundation model, while prior passing points and incident angles are handled separately. Extracted features are combined and fed into a transformer network, which predicts the final passing point and incident angle from distinct output layers.} 
    \vspace{-0.45cm}
    \label{fig:network}
\end{figure}

\begin{figure}
	\centering 
	\includegraphics[width= 0.47\textwidth]{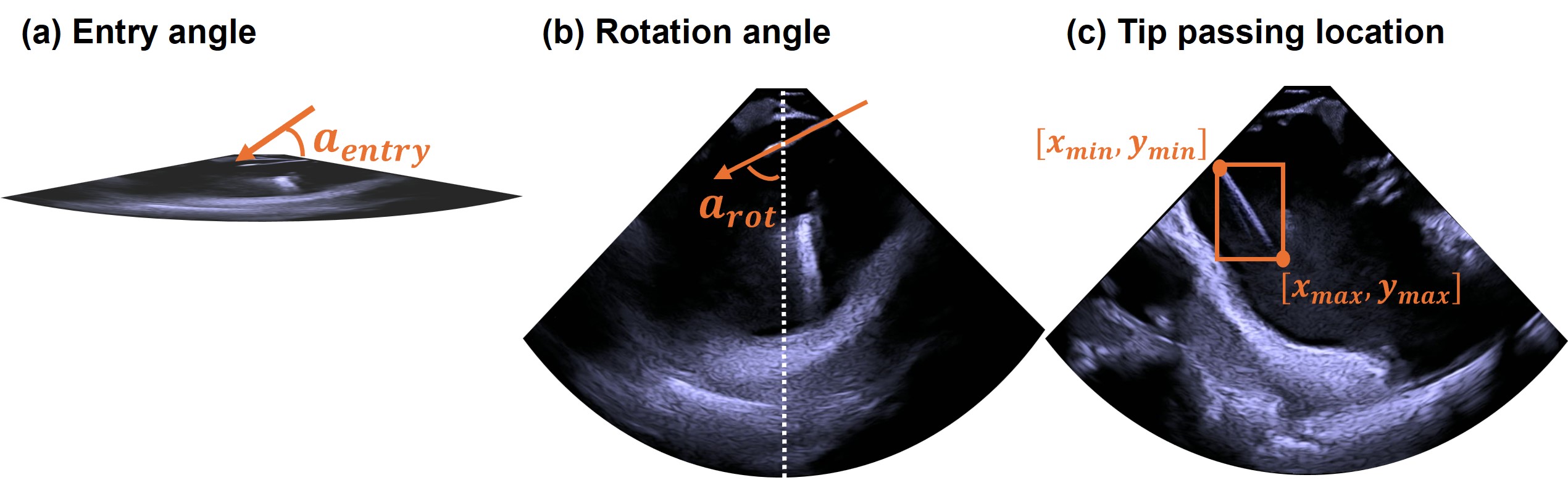}
	\caption{The dataset composition. (a) Entry angle $a_{entry}$: The angle at which the tip enters the US fan area. (b) Rotation angle $a_{rot}$: The rotational angle between the tip and the center-line of the US image in 2D. (C) Tip passing location: The position of the device within the 2D US image.}
    \vspace{-0.45cm}
    \label{fig:dataset}
\end{figure}

\vspace{-0.2cm}
 We propose a method to estimate the passing point of the therapy device within the 2D ICE image plane along with its incident angle, as illustrated in Figure\,\ref{fig:dataset}. Our approach integrates a pretrained ultrasound (US) foundation model and a transformer network. leveraging a 37.4M echocardiography US dataset trained using self-supervised learning\,\cite{amadou2024echoapex} ( Figure\,\ref{fig:network}).

\subsection{Training details}
\vspace{-0.2cm}
The passing point is represented as a bounding box as $B=[x_{min},y_{min},x_{max},y_{max}]$, where $(x_{min},y_{min})$ and $(x_{max},y_{max})$ are the top-left and bottom-right coordinates. The incident angle is defined as $A=[a_{entry},a_{rot}]$, where $a_{entry}$ is the entry angle into the 2D ICE image plane, and $a_{rot}$ denotes its rotational orientation. 
 
A sequence of ICE images $I_{1:N}=[i_{1},i_{2}, ...,i_{N}]$ (N=5 in this work) is processed, resized to $224 \times 224$, and passed through the US foundation model ($M_{foundation}$) to extract feature representations $F_{I}=[f_{1},f_{2},...,f_{N}]$. To ensure temporal consistency, prior passing point $B_{N-1}$ and the incident angle $A_{N-1}$ are projected into the same feature space via a linear transformation.

The extracted features are concatenated with the \textit{[CLS]} token and passed through a transformer network ($M_{main}$) consisting of 8 encoder layers and 6 attention heads. The \textit{[CLS]} output is processed via linear layers to predict the passing point $\hat{B}$ and incident angle $\hat{A}$:
\begin{eqnarray}
    \hat{B}, \hat{A} = M_{main}([CLS,F_{I},B_{N-1},A_{N-1}]),\\
    l_{total} = l_{mse}(\hat{B}, B_N) + l_{mse}(\hat{A}, A_N).
\end{eqnarray}
where $B_N$ and $T_N$ are ground-truth values, and $l_{mse}$ is the MSE loss function.
The model was trained for 117 epochs with a batch size of 6 in PyTorch on a single NVIDIA A100 GPU, achieving real-time performance at 25 [Hz].

\subsection{Dataset}
\vspace{-0.2cm}
Acquiring a large-scale ICE dataset with accurate catheter tip locations and ground-truth incident angles is challenging. To address this, we use recorded clinical sequences and synthetic data generation to simulate various device orientations, incident angles, and anatomical interactions, ensuring clinical relevance.

To capture the catheter tip’s position and orientation, we equipped both the ICE catheter and device catheter tip (9-Fr) with EM sensors, initializing the sensor frame such that ICE catheter and tip heading aligns with the z-axis and ICE US fan direction aligns with the EM x-axis.

ICE images were collected in a water chamber against a black background, enabling automated annotation using computer vision methods. The $a_{rot}$ was inferred from the bounding box diagonal orientation, while the $a_{entry}$ was computed using EM sensor data by $E_{ice}^{tip} = (E_{world}^{ice})^{-1}E_{world}^{tip}$, where $E_{world}$ is the global frame, and $a_{entry}$ was  derived from the z-axis angle in $E_{ice}^{tip}$.

To enhance realism, extracted tip images were overlaid onto real ICE images from clinical ablation procedures, preserving motion continuity and introducing intensity variations for improved generalization. The final dataset consists of 5,698 ICE-tip image pairs, divided into 5,400 training, 48 validation, and 250 test cases. Each case includes sequential frames, ensuring continuous tip movement, essential for real-time tracking model training. One important thing to note is that there is no overlap between the real ICE images and tip sequence images in the training and test sets.

\section*{RESULTS}
\begin{figure}
	\centering 
	\includegraphics[width= 0.47\textwidth]{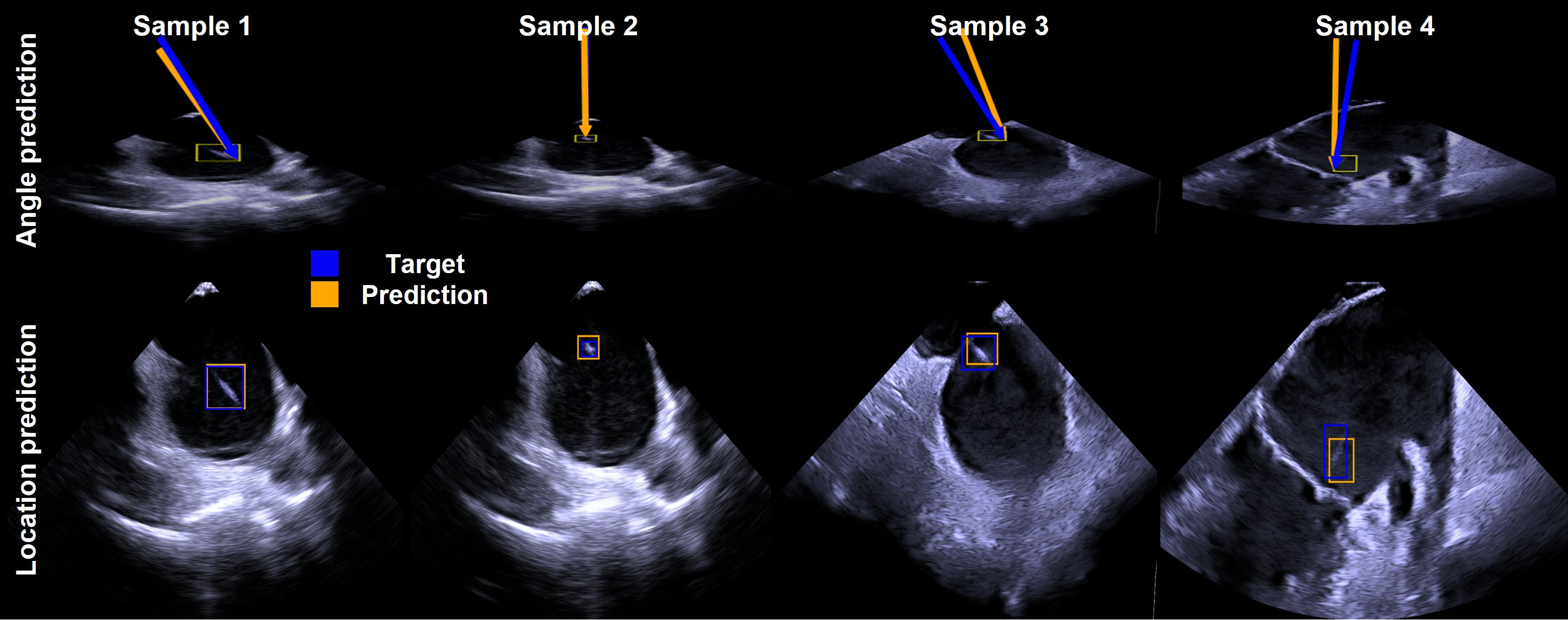}
	\caption{Representative results of the proposed method. Each column presents a different test sample. The first row shows the angular prediction results, while the second row visualizes the tip location prediction. The blue color represents the target tip, whereas the orange color denotes the predicted tip.}
    \vspace{-0.45cm}
    \label{fig:result}
\end{figure}

We evaluated the model using 250 test cases, including 12 distinct motion sequences incorporating variations in insertion, withdrawal, and speed changes (10–20 mm/s). Both qualitative and quantitative validation were performed.
Qualitative Validation: The predicted tip position was visualized, showing strong alignment between the predicted (orange) and target (blue) tips. The predicted bounding box closely overlapped with the target, indicating minimal errors in incident angles (Figure \ref{fig:result}).
Quantitative Validation: 
Predicting the current state from previous estimates across 12 sequences, we achieved an average entry angle error of $13.76\degree \pm 9.58\degree$ and a rotation angle error of $22.99\degree \pm 15.12\degree$, demonstrating high accuracy across motion variations.
Tip Location Accuracy: The Intersection-over-Union (IoU) score averaged 0.66, confirming the model's effectiveness in tracking the therapy device tip under different motion conditions.




\section*{DISCUSSION}
In this study, we developed an AI-driven model to estimate the passing point and incident angle of a therapy device in ICE images, enabling continuous monitoring during ICE-guided interventions. By leveraging a pretrained US foundation model, we extracted spatial and temporal features, while incorporating historical passing points and angular information to improve prediction accuracy and consistency. Evaluation on 250 test cases demonstrated strong alignment between the predicted and target tip positions, with minimal errors in estimated angles.

Integrating this algorithm with robot-assisted ICE control could enable real-time probe adjustments, ensuring continuous tip visibility. While synthetic data improved diversity, expanding with more clinical cases is essential for better generalization and robustness in complex surgical settings.

\section*{DISCLAIMER}
{
The concepts and information presented in this paper are based on research results that are not commercially available. Future availability cannot be guaranteed.
}

\section*{ACKNOWLEDGEMENT}
Research reported in this publication was supported in part by the National Institute of Biomedical Imaging and Bioengineering of the National Institutes of Health under award number R01EB028278. The content is solely the responsibility of the authors and does not necessarily represent the official views of the National Institutes of Health.

{
\bibliography{references_hamlyn}
}

\end{document}